\documentclass{pasj01}
\usepackage{natbib} 
\usepackage{url}
\usepackage{xcolor}

\newcommand{\bvtheta}{\boldsymbol{\vartheta}}
\newcommand{\bd}{\boldsymbol{d}}
\newcommand{\cL}{\mathcal{L}}

\Received{$\langle$reception date$\rangle$}
\Accepted{$\langle$acception date$\rangle$}
\Published{$\langle$publication date$\rangle$}

\begin{document}

\title{New constraints on the mass bias of galaxy clusters from the power spectra of the thermal Sunyaev-Zeldovich effect and cosmic shear}
\author{Ryu \textsc{Makiya}\altaffilmark{1}}
\author{Chiaki \textsc{Hikage}\altaffilmark{1}}
\author{Eiichiro \textsc{Komatsu}\altaffilmark{2,1}}

\altaffiltext{1}{Kavli Institute for the Physics and Mathematics of the Universe (Kavli IPMU, WPI), University
of Tokyo, Chiba 277-8582, Japan}
\altaffiltext{2}{Max-Planck-Institut f\"{u}r Astrophysik, Karl-Schwarzschild Str. 1, 85741 Garching, Germany}

\email{ryu.makiya@ipmu.jp}

\KeyWords{galaxies: clusters: general --- galaxies: clusters: intracluster medium --- cosmology: observations --- large-scale structure of universe}

\maketitle

\begin{abstract}
Thermal Sunyaev-Zeldovich (tSZ) power spectrum is a powerful probe of the present-day amplitude of matter density fluctuations, and has been measured up to $\ell\approx 10^3$ from the {\it Planck} data. 
The largest systematic uncertainty in the interpretation of this data is the so-called ``mass bias'' parameter $B$, which relates the true halo mass to the mass proxy used by the {\it Planck} team as $M_{\rm 500c}^{Planck}=M_{\rm 500c}^{\rm true}/B$.  Since the power spectrum of the cosmic weak lensing shear is also sensitive to the amplitude of matter density fluctuations via $S_8\equiv \sigma_8 \Omega_m^{\alpha}$ with $\alpha\sim 0.5$, we can break the degeneracy between the mass bias and the cosmological parameters by combining the tSZ and cosmic shear power spectra. In this paper,
we perform a joint likelihood analysis of the tSZ power spectrum from {\it Planck} and the cosmic shear power spectrum from Subaru Hyper Suprime-Cam. Our analysis does not use the primordial cosmic microwave background (CMB) information. We obtain a new constraint on the mass bias as $B = 1.37 ^{+0.15}_{-0.23}$ or $(1-b) = B^{-1}=0.73^{+0.08}_{-0.13}$ (68\%~C.L.), for $\sigma_8 < 0.9$.  This value of $B$ is lower than that needed to reconcile the tSZ data with the primordial CMB and CMB lensing data, i.e., $B = 1.64 \pm 0.19$, but is consistent with the mass bias expected from hydrodynamical simulations, $B = 1.28 \pm 0.20$.
Our results thus indicate that the mass bias is consistent with the non-thermal pressure support from mass accretion of galaxy clusters via the cosmic structure formation, and that the cosmologies inferred from the tSZ and the cosmic shear are consistent with each other.
\end{abstract}

\section{Introduction}
\label{sec:introduction}
The thermal Sunyaev-Zeldovich (tSZ) effect (\citealt{sunyaev/zeldovich:1972}) is the spectral distortion of the cosmic microwave background (CMB) through the inverse Compton scattering of CMB photons by hot thermal electrons in galaxy clusters.
The tSZ angular power spectrum is sensitive to the amplitude of matter density fluctuations characterized by $\sigma_8$ and $\Omega_m$ (\citealt{komatsu/kitayama:1999, komatsu/seljak:2002}). 
These parameters are, however, strongly degenerate with the so-called ``mass bias'' parameter $B$, which is defined as the ratio of the mass proxy used by the {\it Planck} team and the true mass of galaxy clusters,
\begin{equation}
  B \equiv M_{\rm 500c}^{\rm true}/M_{\rm 500c}^{\it Planck},
\end{equation}
where $M_{\rm 500c}$ is the mass enclosed by the radius $r_{\rm 500c}$ within which the average mass density is 500 times of the critical density of the Universe.
The mass bias $B$ is related to the more commonly used
parameter $b$ as $B = (1-b)^{-1}$ (\citealt{planck2015_sz/etal:2016}). 
Specifically, the tSZ power spectrum depends primarily on $F \equiv \sigma_8(\Omega_m/B)^{0.40}h^{-0.21}$ \citep{bolliet/etal:2018}, where $\Omega_m$ is the matter density parameter, $\sigma_8$ the r.m.s. matter density fluctuation smoothed over a $8~h^{-1}~{\rm Mpc}$ sphere, and $h$ the dimensionless Hubble constant defined by $H_0=100~h~{\rm km~s^{-1}~Mpc^{-1}}$.

The mass bias was introduced to account for the cluster mass uncertainty in the {\it Planck} analysis. The galaxy cluster masses used by the {\it Planck} team were calibrated against a local cluster sample observed by {\it XMM-Newton}, assuming the hydrostatic equilibrium (HSE) with thermal pressure (\citealt{arnaud/etal:2010}).
The {\it Planck} team reported that 20--40\% mass bias (i.e., $1.25 < B < 1.67$) is required to reconcile the power spectra of tSZ clusters with the joint result of {\it Planck} primordial CMB, CMB lensing and Baryonic Acoustic Oscillations (BAO) (\citealt{planck2015_sz/etal:2016}).
Several authors performed follow-up analyses of the {\it Planck} tSZ power spectrum and found similar results  (\citealt{horowitz/seljak:2017, hurier/lacasa:2017, salvati/etal:2018, bolliet/etal:2018, makiya/etal:2018, salvati/etal:2019}).

On the other hand, state-of-the-art cosmological hydrodynamic simulations show that the HSE mass underestimates the true mass due to non-thermal pressure support (e.g., \citealt{kay/etal:2004, rasia/etal:2006, rasia/etal:2012, nagai/vikhlinin/kravtsov:2007, piffaretti/etal:2008, lau/etal:2009, maneghetti/etal:2010, lau/etal:2013, nelson/etal:2014a, nelson/etal:2014b}). The dominant contribution to the non-thermal pressure support seen in the simulations is the mass accretion of galaxy clusters via structure formation \citep{shi/komatsu:2014, shi/etal:2015}, which yields $B = 1.28 \pm 0.20$ (68\%~C.L.)
over a wide range of dynamical states of galaxy clusters \citep[see Table 1 of ][for the mass-limited sample and the fitting range of $(0.1, 1.5)r^{\rm true}_{500}$]{shi/etal:2016}.
The predicted HSE mass bias is therefore lower than, or at least on the lower side of, the {\it Planck} inferred value. This indicates that other sources of non-thermal pressure (such as cosmic rays and magnetic fields) are larger than expected; there might be other systematic effects such as the calibration error in X-ray observations; and/or that new physics beyond the standard $\Lambda$ Cold Dark Matter (CDM) model, e.g., dark energy different from the cosmological constant \citep{bolliet/etal:2018}, and/or modified gravity, is required to resolve a tension between cosmological parameters inferred from the tSZ clusters and those from primordial CMB based on $\Lambda$CDM \citep{planck2015_sz/etal:2016}.
Massive neutrinos, which also modify the evolution of matter density fluctuations, have been shown not to help resolve this tension \citep{salvati/etal:2018, sakr/etal:2018, bolliet/etal:2019}.

To gain more insights into this potential tension, we need an estimate of the mass bias that is independent of the primordial CMB. To this end, in this paper we perform a joint analysis of power spectra of the late-time Universe probes: tSZ and the cosmic weak lensing shear, to obtain a new constraint on the mass bias that is independent of the primordial CMB.
Cosmic shear is a unique probe of the growth of total matter distribution including dark matter. The angular power spectrum of the cosmic shear is  sensitive to $S_8 \equiv \sigma_8 (\Omega_m/0.3)^{\alpha}$ with $\alpha \sim 0.5$.
This information is useful to break the degeneracy between the cosmological parameters and $B$.  

Here we use the cosmic shear measurements from the Hyper Suprime-Cam Subaru Strategic Program (HSC-SSP, hereafter HSC)      (\citealt{hsc1, hsc2}). 
HSC is a wide-field  imaging survey using a 1.5 deg diameter field-of-view camera mounted on the prime focus of the 8.2m Subaru telescope \citep{Miyazaki18}. The unique property of HSC is a combination of the depth ($i\sim 26$) and the excellent image quality (typical $i$-band seeing is $\sim 0.58"$), which enables us to measure cosmic shear signals with unprecedentedly high precision. \cite{hikage/etal:2019} measured tomographic lensing power spectra using the first-year shear catalog \citep{Mandelbaum18, Oguri18} where the total sky coverage is 137 deg$^2$ and the effective number density is 17 arcmin$^{-2}$ in the range of photometric redshifts from 0.3 to 1.5. The first-year shear catalog has been made publicly available as a part of the second public data release of HSC \citep{HSC_PDR2}.

Constraints on the mass bias from the tSZ and cosmic shear power spectra can be estimated as follows. 
From the definition of $F$, the scaling of mass bias with other parameters can be written as 
\begin{equation}
    \label{eq:bias_scale}
    B = F^{-2.5}\sigma_8^{2.5} \Omega_m h^{-0.53}.
\end{equation}
Combining the scaling relation of $\sigma_8$ from the cosmic shear, $\sigma_8 = S_8 (0.3/\Omega_m)^\alpha$, we find
\begin{equation}
    B = 0.3 \left(\frac{S_8}{F}\right)^{2.5}
    \left(\frac{\Omega_m}{0.3}\right)^{1-2.5\alpha}h^{-0.53}.
\end{equation}
If we put $F  = 0.460 \pm 0.012$ from  {\it Planck} \citep{bolliet/etal:2018} and $S_8 = 0.780^{+0.030}_{-0.033}$ with $\alpha = 0.5$ from  HSC \citep{hikage/etal:2019}, we obtain
\begin{equation}
    \label{eq:bias_om}
    B = (1.36 \pm 0.17)\left(\frac{0.3}{\Omega_m}\right)^{0.25}\left(\frac{0.7}{h}\right)^{0.53}.
\end{equation}
Similarly we can replace $\Omega_m$ in Eq.~(\ref{eq:bias_scale}) with $\sigma_8$. In this case we obtain
\begin{equation}
    \label{eq:bias_sigma8}
    B = (1.37 \pm 0.15)\left(\frac{\sigma_8}{0.8}\right)^{0.5}\left(\frac{0.7}{h}\right)^{0.53}.
\end{equation}
These quick estimates are closer to $B=1.28\pm 0.20$ expected from non-thermal pressure due to mass accretion \citep{shi/etal:2016}, which motivates our calculating $B$ from a joint analysis of the tSZ and cosmic shear power spectra in this paper.

This paper is organized as follows.
In Section \ref{sec:data} we describe the datasets of tSZ and cosmic shear that we use. In Section \ref{sec:analysis} we outline the model of the tSZ power spectrum and the details of our likelihood analysis.
We discuss our results in Section \ref{sec:results} and conclude in Section \ref{sec:summary}.
Throughout the paper, we adopt a flat $\Lambda$CDM cosmology with the minimal total neutrino mass of $\sum m_{\nu} = 0.06\;{\rm eV}$.

\section{Data}
\label{sec:data}
\subsection{tSZ}
\label{sec:tszmap}
We use the tSZ power spectrum data before marginalizing over the foreground components. Specifically, we take the data from Table 3 of \cite{bolliet/etal:2018}, which are based on \cite{planck2015_sz/etal:2016}.

The tSZ power spectrum was calculated by cross-correlating the first-half data of the Needlet Internal Linear Combination (NILC) map and last-half of the Modified Internal Linear Combination Algorithm (MILCA) map, where NILC and MILCA are two different methods for reconstructing the tSZ map.\footnote{In \cite{makiya/etal:2018} we used the cross-power spectrum of  NILC's fist- and last-half maps as the Compton-Y auto spectrum, in order to be consistent with the galaxy--Compton-Y cross spectrum analysis. In this paper we decided to follow \cite{planck2015_sz/etal:2016} and \cite{bolliet/etal:2018} for simplicity.}
We also take into account contributions from residual contaminating sources such as the cosmic infrared background (CIB), IR and radio point sources, and the correlated noise.
The amplitudes of CIB, IR and radio point source power spectra are treated as free parameters, while their shapes, which are also taken from Table 3 of \cite{bolliet/etal:2018}, are fixed.
The amplitude of correlated noise is fixed to reproduce the tSZ spectrum at $\ell = 2742$, following \cite{bolliet/etal:2018}.

\subsection{Cosmic shear}
We use a sample of the posterior distributions of the cosmological parameters of a flat $\Lambda$CDM model with a fixed minimal total neutrino mass $\sum m_{\nu} = 0.06\;{\rm eV}$ obtained from the HSC lensing power spectra alone \citep{hikage/etal:2019}\footnote{The likelihood chain can be found at \url{http://gfarm.ipmu.jp/~surhud/#hikage}}. The parameters include the five basic cosmological parameters ($\Omega_{b}h^2$, $\Omega_{c}h^2$, $h$, $A_{s}$ and $n_s$) and nine nuisance parameters regarding modelling errors of point spread functions (PSF), shear biases, photo-$z$ errors, and intrinsic alignments. The reionization optical depth $\tau$ is not used in the cosmic-shear-alone analysis.
The range of flat priors of cosmological parameters adopted in the HSC cosmic shear analysis are listed in Table \ref{tb:params}.

\section{Analysis}
\label{sec:analysis}
We model and analyze the tSZ power spectrum using the halo-model-based approach that we have established in \cite{bolliet/etal:2018} and \cite{makiya/etal:2018}, which are based on \cite{komatsu/seljak:2002}.\footnote{While we used the mass function of \cite{bocquet/etal:2016} in \cite{makiya/etal:2018}, we follow \cite{bolliet/etal:2018} to use that of \cite{tinker/etal:2008} in this paper. Since their mass functions are for the overdensity with respect to the mean mass density (rather than the critical density), we use the spline interpolation of the parameters at various overdensities to obtain the mass function for $M_{500c}$. As \cite{tinker/etal:2008} only provide the mass function parameters up to the mean mass overdensity of $\Delta_m$ = 3200, we linearly extrapolate the parameters beyond this bound. We have checked the extrapolation against the fitting function provided by \cite{tinker/etal:2008} and found that the extrapolation method does not have significant effects on our results.} See Section 3.1.1 of \cite{makiya/etal:2018} for details of the implementation.

There are two important changes from the methodology established in these papers. First, we properly take into account the effect of massive neutrinos in the modelling of the power spectrum and mass function of collapsed objects (Section~\ref{sec:mnu}). Second, we improve the method to calculate the likelihood when the covariance matrix includes a non-Gaussian term (Section~\ref{sec:like}).

\subsection{Massive neutrinos}
\label{sec:mnu}
We include the effect of massive neutrinos by following the so-called ``CDM prescription'' \citep{ichiki/takada:2012,costanzi/etal:2013,villaescusa-Navarro/etal:2014,castorina/etal:2014}. The basic idea is to remove the contribution of neutrinos to the mass of collapsed objects (halos) when computing statistics of halos, as neutrino stream out of them.
In the calculation of the dark matter halo mass function,
we modify the matter density $\rho_{m}$ and the matter power spectrum $P(k,z)$ as
\begin{equation}
    \rho_{m} = (\Omega_{c}+\Omega_{b})\rho_{\rm crit},
\end{equation}
and
\begin{equation}
\label{eq:pk_nu}
    P(k,z) = \frac{2\pi^3}{k^3}A_s\left(\frac{k}{k_{*}}\right)^{n_s-1}
    \left(
    \frac{\Omega_c \mathcal{T}_c(k,z)+\Omega_b \mathcal{T}_b(k,z)}{\Omega_c + \Omega_b}
    \right)^2,
\end{equation}
where $A_s, n_s$ and $k_{*}=0.05~{\rm Mpc}^{-1}$ are the amplitude, the spectral index and the pivot scale of the primordial power spectrum, $\Omega_c$ and $\Omega_b$ are the density parameters of CDM and baryons, $\rho_{\rm crit}$ is the critical density of the Universe, and $\mathcal{T}_c$ and $\mathcal{T}_b$ are the transfer functions of CDM and baryons respectively.
We refer the reader to \cite{bolliet/etal:2019} for more details of the effects of massive neutrios on the tSZ power spectrum.

\begin{table*}
  \tbl{Mean and 68\% confidence regions of the model parameters.
The range of flat priors are shown in the second column.}{%
  \begin{tabular}{cccccc}
 \hline & & \multicolumn{2}{c}{no $\sigma_8$ cut-off} & \multicolumn{2}{c}{$\sigma_8 < 0.9$} \\
    & Prior    & Mean &  68\% C.L. & Mean & 68\% C.L.\\
    \hline \hline
		$B$ & [0.5, 3.0] & 1.54 & [1.19, 1.74] & 1.37 & [1.13,1.52] \\
		$A_{\rm CIB}$ & [0,1.0] & 0.40 & [0.19,0.58] & 0.47 & [0.27, 0.68] \\
		$A_{\rm IR}$ & [0,2.5]  & 1.97 & [1.82,2.12] & 2.01 & [1.86,2.16] \\
		$A_{\rm Rad}$ & [0,1.5] & 0.31 & [0.00,0.37] & 0.35 & [0.00,0.42] \\
    \hline
        $\Omega_b h^2$ & [0.019, 0.026] & 0.023 & [0.020, 0.025] & 0.023 & [0.020,0.025] \\
        $\Omega_c h^2$ & [0.03, 0.7] & 0.11 & [0.047, 0.14] & 0.16 & [0.11,0.20] \\
        $h$ & [0.6, 0.9] & 0.77 & [0.70, 0.88] & 0.76 & [0.69,0.87] \\
        $\ln{(10^{10}A_s)}$ & [1.5, 6.0] & 3.54 & [2.51, 4.51] & 2.67 & [2.06,3.18] \\
        $n_s$ & [0.87, 1.07] & 0.96 & [0.89, 1.00] & 0.95 & [0.88,0.99] \\
	\hline
        $S_8 \equiv \sigma_8 (\Omega_m/0.3)^{0.5}$         & -- & 0.79 & [0.76, 0.81] & 0.79 & [0.77,0.81] \\
		$F \equiv \sigma_8 (\Omega_m/B)^{0.4} h^{-0.21}$ & -- & 0.45& [0.44, 0.48] & 0.46 & [0.45,0.48] \\
	\hline
    \end{tabular}}\label{tb:params}
\end{table*}

\subsection{Likelihood analysis}
\label{sec:like}
To quickly perform a joint likelihood analysis we use the  importance sampling technique (\citealt{cosmomc}) as follows. 
We leave a full likelihood analysis as a future work.

First we read a set of cosmological parameters from the likelihood chains of the HSC cosmic shear analysis line by line and calculate the tSZ power spectrum and its likelihood $\cL_{\rm tSZ}$ for the same set of cosmological parameters.
Then we reweigh the chain by multiplying a weight by $\exp(-\ln \cL_{\rm tSZ})$.
Applying this procedure to the entire sample of HSC chains, we obtain the tSZ-shear joint posterior distributions.

Our model has four nuisance parameters: the mass bias $B$ and the amplitudes of the power spectra of CIB, IR and radio point sources, $A_{\rm CIB}, A_{\rm IR}$ and $A_{\rm Rad}$, respectively.
The nuisance parameters are randomly picked from a flat prior with the range listed in Table \ref{tb:params}.
Following \cite{bolliet/etal:2018}, the total power of the contaminating sources are restricted not to exceed the residual of the total tSZ power and the sum of the contributions from resolved clusters.
We iterate the above procedure by changing the random seed for nuisance parameters until the chains are converged, and then combine multiple chains to obtain the posterior distributions.
We judge that the chains converge when the Gelman-Rubin estimator $R-1$, where $R$ is the ratio of the variance between chains and within chains (\citealt{gelma/rubin:1992}), is less than 0.05.

The tSZ likelihood is calculated as
\begin{equation}
\label{eq:like}
-2 \ln \cL_{\rm tSZ}(\bd|\bvtheta) = {\Delta}^{\rm T} {\rm Cov}^{-1} \Delta,
\end{equation}
where $\Delta$ is the difference vector between the observed and the model tSZ spectra, and ${\rm Cov}$ is the covariance matrix including the non-Gaussian term calculated from the model tSZ trispectrum.
The Gaussian term of the covariance matrix is taken from Table 3 of \cite{bolliet/etal:2018}. 

In \cite{makiya/etal:2018} we calculated the non-Gaussian term at each step of parameter inference.
However \cite{carron:2013} pointed out that such a parameter-dependent covariance matrix adds extra artificial information and can bias the parameter constraints.
Thus we adopt the new procedure in this paper.
First, we perform a likelihood analysis without the non-Gaussian term and find the best-fitting parameters.
Then we calculate the non-Gaussian term from the best-fitting parameters and repeat the likelihood analysis by including the fixed non-Gaussian term in the covariance matrix. From the new best-fitting parameters, we recalculate the covariance matrix and redetermine the best-fitting parameters. We iterate this procedure until the best-fitting parameters converge.

\section{Results}
\label{sec:results}
\begin{figure*}[htbp]
 \includegraphics[width=2\columnwidth, bb=0 0 840 853]{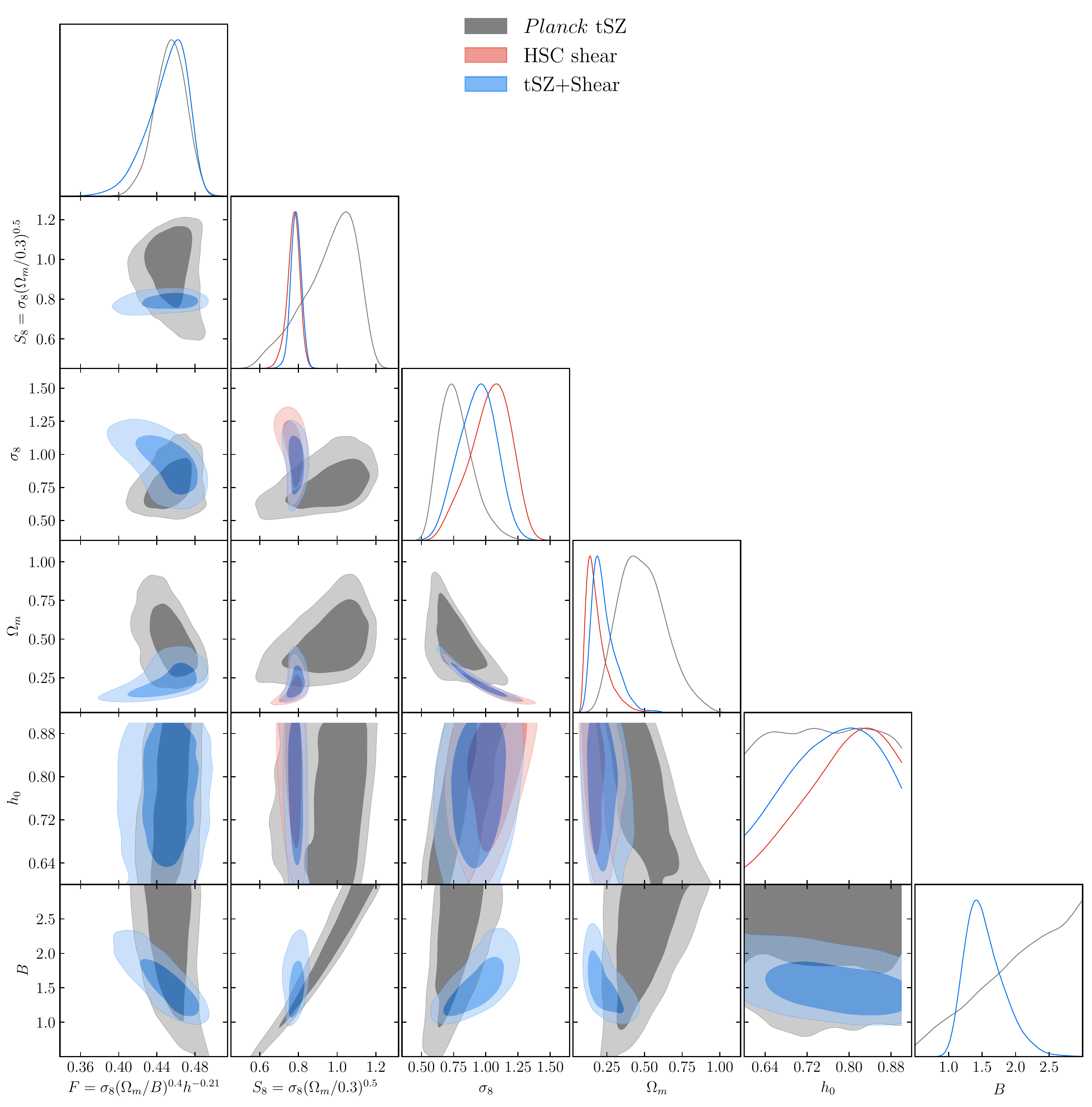}
 \caption{
 The 1D and 2D posterior distributions of $\sigma_8$, $\Omega_m$, $h$ and $B$, marginalized over the other parameters, inferred from the analysis of the {\it Planck} tSZ alone (gray), the HSC cosmic shear alone (red) and the joint analysis of them (blue).
 The 2D contours show the 68\% and 95\% confidence levels. The contours are smoothed by a Gaussian of 0.3 times standard deviations in each parameter for  clarity. The smoothing does not affect the statistical analysis of the parameters. }
 \label{fig:params_tri_alone}
\end{figure*}

Figure \ref{fig:params_tri_alone} shows the constraints on the parameters from the {\it Planck} tSZ alone, the HSC shear alone, and the joint analysis of them. 
The results from the joint likelihood analysis are summarized in Table \ref{tb:params}.
The nuisance parameters not shown in Figure \ref{fig:params_tri_alone} ( $A_{\rm CIB}$, $A_{\rm IR}$, $A_{\rm Rad}$) are well determined within the prior ranges, as shown in Table \ref{tb:params}.
The constraints on $S_8$ and $F$ are consistent with those obtained from the HSC cosmic shear and the {\it Planck} tSZ power spectrum alone, $S_8 = 0.780^{+0.030}_{-0.033}$ \citep{hikage/etal:2019} and $F = 0.460 \pm 0.012$ \citep{bolliet/etal:2018}, respectively.
Since the cosmic shear and tSZ power spectra are not sensitive to individual parameters (i.e., all cosmological parameters and mass bias parameter) but only sensitive to the combination of them, $S_8$ (for the cosmic shear) and $F$ (for the tSZ), 
the best-fitting model from the joint analysis also gives a good fit to both data sets.

As shown in Figure \ref{fig:params_tri_alone}, the mass bias cannot be determined from the tSZ data alone because of the degeneracy with $\sigma_8$, $\Omega_m$ and $h$. The HSC cosmic shear lifts this degeneracy and helps to determine the mass bias.
We find that the constraints from the joint analysis do not fully overlap with those from the tSZ alone or the shear alone analysis.
This is due to a large difference between the best-fitting values of the tSZ alone or shear alone analysis and those from the joint analysis. The contour shows the likelihood distance from the best-fitting point. When the best-fitting values of two different distributions are largely different, direct comparison of two contours is not meaningful because those contours measure the likelihood distances from different points.

\begin{figure*}
 \includegraphics[width=2\columnwidth, bb=0 0 576 252]{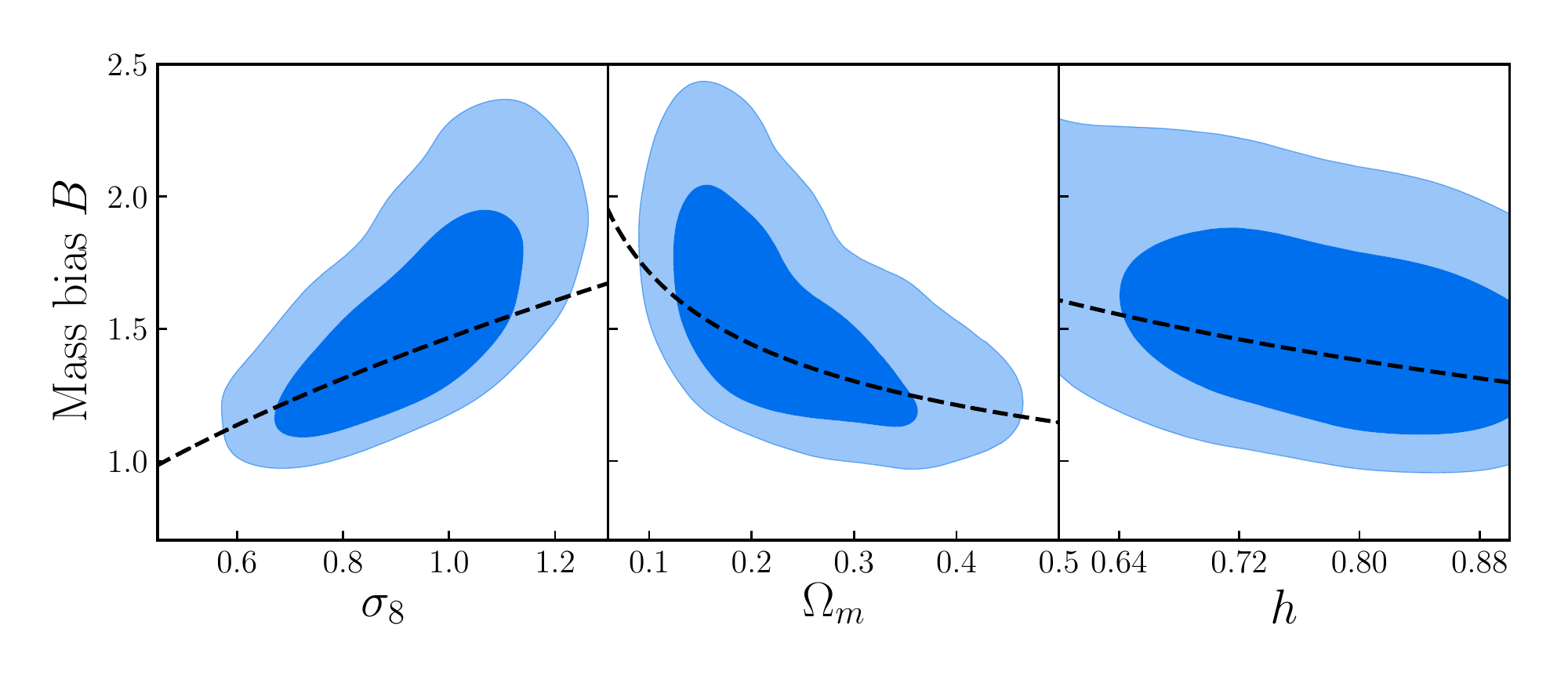}
 \caption{
 The 2D posterior distributions of $\sigma_8$, $\Omega_m$, $h$ and $B$, marginalized over the other parameters, inferred from the joint analysis of the {\it Planck} tSZ and HSC cosmic shear power spectra.
 The contours show the 68\% and 95\% confidence levels. The contours are smoothed by a Gaussian of 0.3 times standard deviations in each parameter for clarity.
 The dashed lines show the predicted scaling relations of the mass bias and the other parameters (see Section \ref{sec:introduction} and \ref{sec:results} for details).}
 \label{fig:params_tri}
\end{figure*}

We find that the {\it Planck} tSZ and the HSC cosmic shear power spectra constrain the mass bias as $B = 1.54^{+0.20}_{-0.35}$ (mean and 68\% C.L.).
The black lines in Figure \ref{fig:params_tri} are the predicted scaling relation shown in Eq.~(\ref{eq:bias_om}) and (\ref{eq:bias_sigma8}) with $\alpha = 0.5$. We used $h = 0.76$ for $B$--$\Omega_m$ and $B$--$\sigma_8$ relations and $\Omega_m = 0.23$ for $B$--$h$ relation, which are the mean value from the joint analysis, respectively. The measurements are consistent with the predicted relations.

Figure \ref{fig:params_tri} shows that the Hubble parameter $h$ is not well constrained within the range of prior.
While the limited range of the prior of $h$ has an effect on the constraint on $B$, the effect is not significant since  dependence of $B$ on $h$ is weak.
We have also examined a Gaussian prior of $h = 0.74 \pm 0.014$ taken from the distance ladder method \citep{riess/etal:2019} and found that the constraint on $B$ does not improve significantly; $B = 1.60^{+0.21}_{-0.33}$. This is again due to the weak $h$ dependence of $B$.

\begin{figure}
\includegraphics[width=\columnwidth, bb=0 0 216 216]{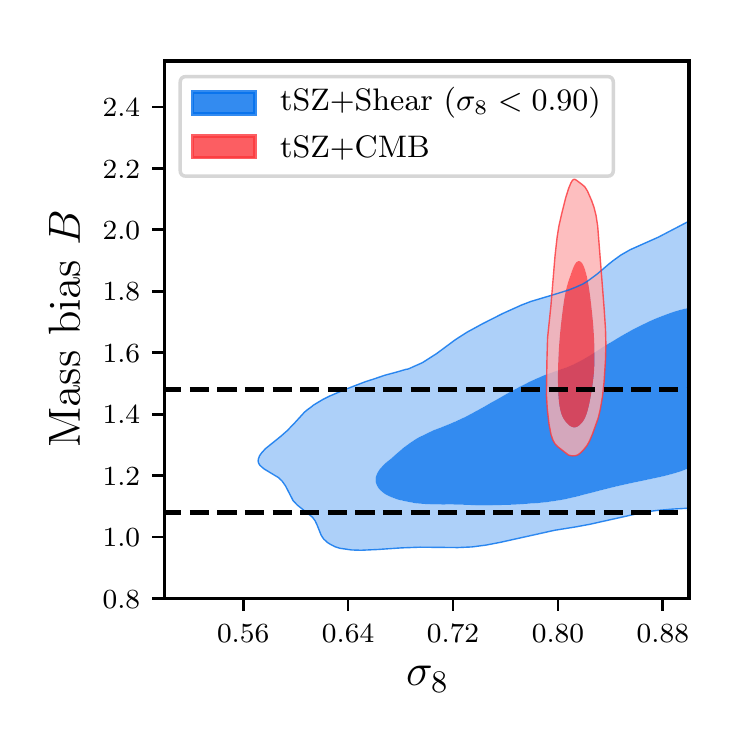}
\caption{Marginalized constraints on $\sigma_8$ and the mass bias $B$ from the joint analysis of the {\it Planck} tSZ and HSC cosmic shear power spectra (blue), and those from the {\it Planck} tSZ, primordial CMB and CMB lensing power spectra (red). 
The tSZ$+$shear contours are the same as those in the lower middle panel of Figure~\ref{fig:params_tri_alone} except a prior of $\sigma_8 < 0.9$.
The contours show the 68\% and 95\% confidence levels and are smoothed by a Gaussian of 0.3 times standard deviations in each parameter. 
The horizontal dashed lines show the 68\% confidence region of the mass bias estimated from the hydrodynamical simulations, $B = 1.28 \pm 0.20$ (\citealt{shi/etal:2016}). 
}
\label{fig:params2}
\end{figure}

Constraint on the mass bias from our analysis is mainly limited by the uncertainty of $\sigma_8$. Currently, most (if not all) data sets indicate and are consistent with $\sigma_8 < 0.9$ \citep{planck2015_cosmo/etal:2016,alam/etal:2017,DESyr1:2018,burenin:2018}. 
To incorporate this knowledge into our analysis, in Figure \ref{fig:params2} we show marginalized joint constraints on $B$ and $\sigma_8$ with a prior of $\sigma_8 < 0.9$. 
This $\sigma_8$ cut roughly corresponds to $\Omega_m > 0.2$ (see the $\sigma_8$--$\Omega_m$ panel of Figure \ref{fig:params_tri_alone}), which is also consistent with most (if not all) data sets.
In this case the mass bias is constrained to be $B = 1.37 ^{+0.15}_{-0.23}$ or $(1-b) = B^{-1}=0.73^{+0.08}_{-0.13}$ (mean and 68\% C.L.). 
This is lower than that from, but is still consistent at 68\% C.L. with, the joint analysis of the tSZ and CMB including CMB lensing, $B = 1.64 \pm 0.19$.\footnote{This constraint is different from that shown in \cite{makiya/etal:2018} due to the difference of dark matter mass function, the tSZ auto-power spectrum data and the treatment of the non-Gaussian term of the covariance matrix, as noted in Section \ref{sec:data} and \ref{sec:analysis}.} 
On the other hand, our result is consistent with the value expected from cosmological hydrodynamical simulations, $B = 1.28 \pm 0.20$ (\citealt{shi/etal:2016}) and also with that estimated from weak lensing mass, $B = 1.25 \pm 0.13$ (\citealt{salvati/etal:2018} and references therein; see also \citealt{miyatake/etal:2019,stern/etal:2019,dietrich/etal:2019} for recent attempts).

\section{Conclusion}
\label{sec:summary}
In this paper we have performed a joint likelihood analysis of the tSZ angular power spectrum obtained by {\it Planck} and the cosmic shear angular power spectra obtained by Subaru HSC.
We have found that the mass bias of {\it Planck} clusters is constrained to be $B = 1.37 ^{+0.15}_{-0.23}$ or $(1-b) = B^{-1}=0.73^{+0.08}_{-0.13}$ (mean and $68\%$ C.L.) for $\sigma_8 < 0.9$.

Our result is consistent with the HSE mass bias estimated from the cosmological hydrodynamical simulations, which predict $B = 1.28 \pm 0.20$ over a wide range of dynamical states of galaxy clusters \citep{shi/etal:2016}. The origin of this bias is non-thermal motion arising from mass accretion of galaxy clusters via structure formation \citep{shi/komatsu:2014,shi/etal:2015}.
Therefore, as long as we adopt the cosmological parameters inferred from the HSC cosmic shear with a weak prior of $\sigma_8<0.9$, the origin of the mass bias can be mostly understood. 
In other words, the cosmological parameters inferred from two different probes of a late-time Universe, tSZ and cosmic shear, are consistent with each other when the mass bias agrees with the expectations from cosmological hydrodynamical simulations.
It has been known that the cosmological parameters inferred from the primordial CMB and those from the late-time probes are in a mild tension (e.g., \citealt{riess/etal:2018b, riess/etal:2018a}, \citealt{joudaki/etal:2017, kohlinger/etal:2017, troxel/etal:2018, burenin:2018,hikage/etal:2019}).
Our results suggest that the tSZ power spectrum may also be in tension with the primordial CMB; a higher value of $B$ reported by {\it Planck} may come from the tension of $\sigma_8$ (or $S_8$) between {\it Planck} and the late-time Universe probes of the cosmic shear and SZ clusters.

More accurate measurements of cosmic shear is required to obtain a robust conclusion on this issue. In this paper we have used the HSC year 1 data, which are based on only 11$\%$ of the planned HSC survey data. Full HSC survey will put tighter constraints on $S_8$ and also improve a constraint on the mass bias. Combining other probes such as galaxy-galaxy lensing and galaxy clustering will improve the constraints.

\begin{ack}
R.~M. thanks Boris Bolliet for useful discussion.
Kavli IPMU is supported by World Premier International Research Center Initiative (WPI), MEXT, Japan.
This work was supported in part by JSPS KAKENHI Grant Number JP15H05896, JP15K21733, JP16K17684 and JP18H04348.

The Hyper Suprime-Cam (HSC) collaboration includes the astronomical communities of Japan and Taiwan, and Princeton University. The HSC instrumentation and software were developed by the National Astronomical Observatory of Japan (NAOJ), the Kavli Institute for the Physics and Mathematics of the Universe (Kavli IPMU), the University of Tokyo, the High Energy Accelerator Research Organization (KEK), the Academia Sinica Institute for Astronomy and Astrophysics in Taiwan (ASIAA), and Princeton University. Funding was contributed by the FIRST program from Japanese Cabinet Office, the Ministry of Education, Culture, Sports, Science and Technology (MEXT), the Japan Society for the Promotion of Science (JSPS), Japan Science and Technology Agency (JST), the Toray Science Foundation, NAOJ, Kavli IPMU, KEK, ASIAA, and Princeton University. 

This paper makes use of software developed for the Large Synoptic Survey Telescope. We thank the LSST Project for making their code available as free software at  http://dm.lsst.org

The Pan-STARRS1 Surveys (PS1) have been made possible through contributions of the Institute for Astronomy, the University of Hawaii, the Pan-STARRS Project Office, the Max-Planck Society and its participating institutes, the Max Planck Institute for Astronomy, Heidelberg and the Max Planck Institute for Extraterrestrial Physics, Garching, The Johns Hopkins University, Durham University, the University of Edinburgh, Queen’s University Belfast, the Harvard-Smithsonian Center for Astrophysics, the Las Cumbres Observatory Global Telescope Network Incorporated, the National Central University of Taiwan, the Space Telescope Science Institute, the National Aeronautics and Space Administration under Grant No. NNX08AR22G issued through the Planetary Science Division of the NASA Science Mission Directorate, the National Science Foundation under Grant No. AST-1238877, the University of Maryland, and Eotvos Lorand University (ELTE) and the Los Alamos National Laboratory.

Based [in part] on data collected at the Subaru Telescope and retrieved from the HSC data archive system, which is operated by Subaru Telescope and Astronomy Data Center at National Astronomical Observatory of Japan.
\end{ack}

\bibliographystyle{apj}
\bibliography{references}

\begin{thebibliography}{}
\expandafter\ifx\csname natexlab\endcsname\relax\def\natexlab#1{#1}\fi

\bibitem[{{Abbott} {et~al.}(2018){Abbott}, {Abdalla}, {Alarcon}, {Aleksi{\'c}},
  {Allam}, {Allen}, {Amara}, {Annis}, {Asorey}, and {Avila}}]{DESyr1:2018}
{Abbott}, T.~M.~C., {Abdalla}, F.~B., {Alarcon}, A., {et~al.} 2018, \prd, 98,
  043526

\bibitem[{{Aihara} {et~al.}(2018{\natexlab{a}}){Aihara}, {Armstrong},
  {Bickerton}, {Bosch}, {Coupon}, {Furusawa}, {Hayashi}, {Ikeda}, {Kamata},
  {Karoji}, {Kawanomoto}, {Koike}, {Komiyama}, {Lang}, {Lupton}, {Mineo},
  {Miyatake}, {Miyazaki}, {Morokuma}, {Obuchi}, {Oishi}, {Okura}, {Price},
  {Takata}, {Tanaka}, {Tanaka}, {Tanaka}, {Uchida}, {Uraguchi}, {Utsumi},
  {Wang}, {Yamada}, {Yamanoi}, {Yasuda}, {Arimoto}, {Chiba}, {Finet},
  {Fujimori}, {Fujimoto}, {Furusawa}, {Goto}, {Goulding}, {Gunn}, {Harikane},
  {Hattori}, {Hayashi}, {He{\l}miniak}, {Higuchi}, {Hikage}, {Ho}, {Hsieh},
  {Huang}, {Huang}, {Imanishi}, {Iwata}, {Jaelani}, {Jian}, {Kashikawa},
  {Katayama}, {Kojima}, {Konno}, {Koshida}, {Kusakabe}, {Leauthaud}, {Lee},
  {Lin}, {Lin}, {Mandelbaum}, {Matsuoka}, {Medezinski}, {Miyama}, {Momose},
  {More}, {More}, {Mukae}, {Murata}, {Murayama}, {Nagao}, {Nakata}, {Niida},
  {Niikura}, {Nishizawa}, {Oguri}, {Okabe}, {Ono}, {Onodera}, {Onoue}, {Ouchi},
  {Pyo}, {Shibuya}, {Shimasaku}, {Simet}, {Speagle}, {Spergel}, {Strauss},
  {Sugahara}, {Sugiyama}, {Suto}, {Suzuki}, {Tait}, {Takada}, {Terai}, {Toba},
  {Turner}, {Uchiyama}, {Umetsu}, {Urata}, {Usuda}, {Yeh}, and {Yuma}}]{hsc2}
{Aihara}, H., {Armstrong}, R., {Bickerton}, S., {et~al.} 2018{\natexlab{a}},
  \pasj, 70, S8

\bibitem[{{Aihara} {et~al.}(2018{\natexlab{b}}){Aihara}, {Arimoto},
  {Armstrong}, {Arnouts}, {Bahcall}, {Bickerton}, {Bosch}, {Bundy}, {Capak},
  {Chan}, {Chiba}, {Coupon}, {Egami}, {Enoki}, {Finet}, {Fujimori}, {Fujimoto},
  {Furusawa}, {Furusawa}, {Goto}, {Goulding}, {Greco}, {Greene}, {Gunn},
  {Hamana}, {Harikane}, {Hashimoto}, {Hattori}, {Hayashi}, {Hayashi},
  {He{\l}miniak}, {Higuchi}, {Hikage}, {Ho}, {Hsieh}, {Huang}, {Huang},
  {Ikeda}, {Imanishi}, {Inoue}, {Iwasawa}, {Iwata}, {Jaelani}, {Jian},
  {Kamata}, {Karoji}, {Kashikawa}, {Katayama}, {Kawanomoto}, {Kayo}, {Koda},
  {Koike}, {Kojima}, {Komiyama}, {Konno}, {Koshida}, {Koyama}, {Kusakabe},
  {Leauthaud}, {Lee}, {Lin}, {Lin}, {Lupton}, {Mand elbaum}, {Matsuoka},
  {Medezinski}, {Mineo}, {Miyama}, {Miyatake}, {Miyazaki}, {Momose}, {More},
  {More}, {Moritani}, {Moriya}, {Morokuma}, {Mukae}, {Murata}, {Murayama},
  {Nagao}, {Nakata}, {Niida}, {Niikura}, {Nishizawa}, {Obuchi}, {Oguri},
  {Oishi}, {Okabe}, {Okamoto}, {Okura}, {Ono}, {Onodera}, {Onoue}, {Osato},
  {Ouchi}, {Price}, {Pyo}, {Sako}, {Sawicki}, {Shibuya}, {Shimasaku},
  {Shimono}, {Shirasaki}, {Silverman}, {Simet}, {Speagle}, {Spergel},
  {Strauss}, {Sugahara}, {Sugiyama}, {Suto}, {Suyu}, {Suzuki}, {Tait},
  {Takada}, {Takata}, {Tamura}, {Tanaka}, {Tanaka}, {Tanaka}, {Tanaka},
  {Terai}, {Terashima}, {Toba}, {Tominaga}, {Toshikawa}, {Turner}, {Uchida},
  {Uchiyama}, {Umetsu}, {Uraguchi}, {Urata}, {Usuda}, {Utsumi}, {Wang}, {Wang},
  {Wong}, {Yabe}, {Yamada}, {Yamanoi}, {Yasuda}, {Yeh}, {Yonehara}, and
  {Yuma}}]{hsc1}
{Aihara}, H., {Arimoto}, N., {Armstrong}, R., {et~al.} 2018{\natexlab{b}},
  \pasj, 70, S4

\bibitem[{{Aihara} {et~al.}(2019){Aihara}, {AlSayyad}, {Ando}, {Armstrong},
  {Bosch}, {Egami}, {Furusawa}, {Furusawa}, {Goulding}, {Harikane}, {Hikage},
  {Ho}, {Hsieh}, {Huang}, {Ikeda}, {Imanishi}, {Ito}, {Iwata}, {Jaelani},
  {Kakuma}, {Kawana}, {Kikuta}, {Kobayashi}, {Koike}, {Komiyama}, {Li},
  {Liang}, {Lin}, {Luo}, {Lupton}, {MacArthur}, {Matsuoka}, {Mineo},
  {Miyatake}, {Miyazaki}, {More}, {Murata}, {Namiki}, {Nishizawa}, {Oguri},
  {Okabe}, {Okamoto}, {Okura}, {Ono}, {Onodera}, {Onoue}, {Osato}, {Ouchi},
  {Shibuya}, {Strauss}, {Sugiyama}, {Suto}, {Takada}, {Takagi}, {Takata},
  {Takita}, {Tanaka}, {Terai}, {Toba}, {Uchiyama}, {Utsumi}, {Wang}, {Wang},
  and {Yamada}}]{HSC_PDR2}
{Aihara}, H., {AlSayyad}, Y., {Ando}, M., {et~al.} 2019, arXiv e-prints,
  arXiv:1905.12221

\bibitem[{{Alam} {et~al.}(2017){Alam}, {Ata}, {Bailey}, {Beutler}, {Bizyaev},
  {Blazek}, {Bolton}, {Brownstein}, {Burden}, and {Chuang}}]{alam/etal:2017}
{Alam}, S., {Ata}, M., {Bailey}, S., {et~al.} 2017, \mnras, 470, 2617

\bibitem[{{Arnaud} {et~al.}(2010){Arnaud}, {Pratt}, {Piffaretti},
  {B{\"o}hringer}, {Croston}, and {Pointecouteau}}]{arnaud/etal:2010}
{Arnaud}, M., {Pratt}, G.~W., {Piffaretti}, R., {et~al.} 2010, \aap, 517, A92

\bibitem[{{Bocquet} {et~al.}(2016){Bocquet}, {Saro}, {Dolag}, and
  {Mohr}}]{bocquet/etal:2016}
{Bocquet}, S., {Saro}, A., {Dolag}, K., \& {Mohr}, J.~J. 2016, \mnras, 456,
  2361

\bibitem[{{Bolliet} {et~al.}(2019){Bolliet}, {Brinckmann}, {Chluba}, and
  {Lesgourgues}}]{bolliet/etal:2019}
{Bolliet}, B., {Brinckmann}, T., {Chluba}, J., \& {Lesgourgues}, J. 2019, arXiv
  e-prints, arXiv:1906.10359

\bibitem[{{Bolliet} {et~al.}(2018){Bolliet}, {Comis}, {Komatsu}, and
  {Mac{\'{\i}}as-P{\'e}rez}}]{bolliet/etal:2018}
{Bolliet}, B., {Comis}, B., {Komatsu}, E., \& {Mac{\'{\i}}as-P{\'e}rez}, J.~F.
  2018, \mnras, 477, 4957

\bibitem[{{Burenin}(2018)}]{burenin:2018}
{Burenin}, R.~A. 2018, Astronomy Letters, 44, 653

\bibitem[{{Carron}(2013)}]{carron:2013}
{Carron}, J. 2013, \aap, 551, A88

\bibitem[{{Castorina} {et~al.}(2014){Castorina}, {Sefusatti}, {Sheth},
  {Villaescusa-Navarro}, and {Viel}}]{castorina/etal:2014}
{Castorina}, E., {Sefusatti}, E., {Sheth}, R.~K., {Villaescusa-Navarro}, F., \&
  {Viel}, M. 2014, Journal of Cosmology and Astro-Particle Physics, 2014, 049

\bibitem[{{Costanzi} {et~al.}(2013){Costanzi}, {Villaescusa-Navarro}, {Viel},
  {Xia}, {Borgani}, {Castorina}, and {Sefusatti}}]{costanzi/etal:2013}
{Costanzi}, M., {Villaescusa-Navarro}, F., {Viel}, M., {et~al.} 2013, Journal
  of Cosmology and Astro-Particle Physics, 2013, 012

\bibitem[{{Dietrich} {et~al.}(2019){Dietrich}, {Bocquet}, {Schrabback},
  {Applegate}, {Hoekstra}, {Grandis}, {Mohr}, {Allen}, {Bayliss}, {Benson},
  {Bleem}, {Brodwin}, {Bulbul}, {Capasso}, {Chiu}, {Crawford}, {Gonzalez}, {de
  Haan}, {Klein}, {von der Linden}, {Mantz}, {Marrone}, {McDonald},
  {Raghunathan}, {Rapetti}, {Reichardt}, {Saro}, {Stalder}, {Stark}, {Stern},
  and {Stubbs}}]{dietrich/etal:2019}
{Dietrich}, J.~P., {Bocquet}, S., {Schrabback}, T., {et~al.} 2019, \mnras, 483,
  2871

\bibitem[{{Gelman} \& {Rubin}(1992){Gelman} and {Rubin}}]{gelma/rubin:1992}
{Gelman}, A., \& {Rubin}, D.~B. 1992, Statistical Science, 7, 457

\bibitem[{{Hikage} {et~al.}(2019){Hikage}, {Oguri}, {Hamana}, {More},
  {Mandelbaum}, {Takada}, {K{\"o}hlinger}, {Miyatake}, {Nishizawa}, {Aihara},
  {Armstrong}, {Bosch}, {Coupon}, {Ducout}, {Ho}, {Hsieh}, {Komiyama},
  {Lanusse}, {Leauthaud}, {Lupton}, {Medezinski}, {Mineo}, {Miyama},
  {Miyazaki}, {Murata}, {Murayama}, {Shirasaki}, {Sif{\'o}n}, {Simet},
  {Speagle}, {Spergel}, {Strauss}, {Sugiyama}, {Tanaka}, {Utsumi}, {Wang}, and
  {Yamada}}]{hikage/etal:2019}
{Hikage}, C., {Oguri}, M., {Hamana}, T., {et~al.} 2019, \pasj, 71, 43

\bibitem[{{Horowitz} \& {Seljak}(2017){Horowitz} and
  {Seljak}}]{horowitz/seljak:2017}
{Horowitz}, B., \& {Seljak}, U. 2017, \mnras, 469, 394

\bibitem[{{Hurier} \& {Lacasa}(2017){Hurier} and {Lacasa}}]{hurier/lacasa:2017}
{Hurier}, G., \& {Lacasa}, F. 2017, \aap, 604, A71

\bibitem[{{Ichiki} \& {Takada}(2012){Ichiki} and {Takada}}]{ichiki/takada:2012}
{Ichiki}, K., \& {Takada}, M. 2012, \prd, 85, 063521

\bibitem[{{Joudaki} {et~al.}(2017){Joudaki}, {Blake}, {Heymans}, {Choi},
  {Harnois-Deraps}, {Hildebrandt}, {Joachimi}, {Johnson}, {Mead}, and
  {Parkinson}}]{joudaki/etal:2017}
{Joudaki}, S., {Blake}, C., {Heymans}, C., {et~al.} 2017, \mnras, 465, 2033

\bibitem[{{Kay} {et~al.}(2004){Kay}, {Thomas}, {Jenkins}, and
  {Pearce}}]{kay/etal:2004}
{Kay}, S.~T., {Thomas}, P.~A., {Jenkins}, A., \& {Pearce}, F.~R. 2004, \mnras,
  355, 1091

\bibitem[{{K{\"o}hlinger} {et~al.}(2017){K{\"o}hlinger}, {Viola}, {Joachimi},
  {Hoekstra}, {van Uitert}, {Hildebrandt}, {Choi}, {Erben}, {Heymans}, and
  {Joudaki}}]{kohlinger/etal:2017}
{K{\"o}hlinger}, F., {Viola}, M., {Joachimi}, B., {et~al.} 2017, \mnras, 471,
  4412

\bibitem[{{Komatsu} \& {Kitayama}(1999){Komatsu} and
  {Kitayama}}]{komatsu/kitayama:1999}
{Komatsu}, E., \& {Kitayama}, T. 1999, \apjl, 526, L1

\bibitem[{{Komatsu} \& {Seljak}(2002){Komatsu} and
  {Seljak}}]{komatsu/seljak:2002}
{Komatsu}, E., \& {Seljak}, U. 2002, \mnras, 336, 1256

\bibitem[{{Lau} {et~al.}(2009){Lau}, {Kravtsov}, and {Nagai}}]{lau/etal:2009}
{Lau}, E.~T., {Kravtsov}, A.~V., \& {Nagai}, D. 2009, \apj, 705, 1129

\bibitem[{{Lau} {et~al.}(2013){Lau}, {Nagai}, and {Nelson}}]{lau/etal:2013}
{Lau}, E.~T., {Nagai}, D., \& {Nelson}, K. 2013, \apj, 777, 151

\bibitem[{Lewis \& Bridle(2002)Lewis and Bridle}]{cosmomc}
Lewis, A., \& Bridle, S. 2002, Phys. Rev., D66, 103511

\bibitem[{{Makiya} {et~al.}(2018){Makiya}, {Ando}, and
  {Komatsu}}]{makiya/etal:2018}
{Makiya}, R., {Ando}, S., \& {Komatsu}, E. 2018, \mnras, 480, 3928

\bibitem[{{Mandelbaum} {et~al.}(2018){Mandelbaum}, {Miyatake}, {Hamana},
  {Oguri}, {Simet}, {Armstrong}, {Bosch}, {Murata}, {Lanusse}, {Leauthaud},
  {Coupon}, {More}, {Takada}, {Miyazaki}, {Speagle}, {Shirasaki}, {Sif{\'o}n},
  {Huang}, {Nishizawa}, {Medezinski}, {Okura}, {Okabe}, {Czakon}, {Takahashi},
  {Coulton}, {Hikage}, {Komiyama}, {Lupton}, {Strauss}, {Tanaka}, and
  {Utsumi}}]{Mandelbaum18}
{Mandelbaum}, R., {Miyatake}, H., {Hamana}, T., {et~al.} 2018, \pasj, 70, S25

\bibitem[{{Meneghetti} {et~al.}(2010){Meneghetti}, {Rasia}, {Merten},
  {Bellagamba}, {Ettori}, {Mazzotta}, {Dolag}, and
  {Marri}}]{maneghetti/etal:2010}
{Meneghetti}, M., {Rasia}, E., {Merten}, J., {et~al.} 2010, \aap, 514, A93

\bibitem[{{Miyatake} {et~al.}(2019){Miyatake}, {Battaglia}, {Hilton},
  {Medezinski}, {Nishizawa}, {More}, {Aiola}, {Bahcall}, {Bond}, {Calabrese},
  {Choi}, {Devlin}, {Dunkley}, {Dunner}, {Fuzia}, {Gallardo}, {Gralla},
  {Hasselfield}, {Halpern}, {Hikage}, {Hill}, {Hincks}, {Hlo{\v{z}}ek},
  {Huffenberger}, {Hughes}, {Koopman}, {Kosowsky}, {Louis}, {Madhavacheril},
  {McMahon}, {Mandelbaum}, {Marriage}, {Maurin}, {Miyazaki}, {Moodley},
  {Murata}, {Naess}, {Newburgh}, {Niemack}, {Nishimichi}, {Okabe}, {Oguri},
  {Osato}, {Page}, {Partridge}, {Robertson}, {Sehgal}, {Sherwin}, {Shirasaki},
  {Sievers}, {Sif{\'o}n}, {Simon}, {Spergel}, {Staggs}, {Stein}, {Takada},
  {Trac}, {Umetsu}, {van Engelen}, and {Wollack}}]{miyatake/etal:2019}
{Miyatake}, H., {Battaglia}, N., {Hilton}, M., {et~al.} 2019, \apj, 875, 63

\bibitem[{{Miyazaki} {et~al.}(2018){Miyazaki}, {Komiyama}, {Kawanomoto}, {Doi},
  {Furusawa}, {Hamana}, {Hayashi}, {Ikeda}, {Kamata}, {Karoji}, {Koike},
  {Kurakami}, {Miyama}, {Morokuma}, {Nakata}, {Namikawa}, {Nakaya}, {Nariai},
  {Obuchi}, {Oishi}, {Okada}, {Okura}, {Tait}, {Takata}, {Tanaka}, {Tanaka},
  {Terai}, {Tomono}, {Uraguchi}, {Usuda}, {Utsumi}, {Yamada}, {Yamanoi},
  {Aihara}, {Fujimori}, {Mineo}, {Miyatake}, {Oguri}, {Uchida}, {Tanaka},
  {Yasuda}, {Takada}, {Murayama}, {Nishizawa}, {Sugiyama}, {Chiba}, {Futamase},
  {Wang}, {Chen}, {Ho}, {Liaw}, {Chiu}, {Ho}, {Lai}, {Lee}, {Jeng}, {Iwamura},
  {Armstrong}, {Bickerton}, {Bosch}, {Gunn}, {Lupton}, {Loomis}, {Price},
  {Smith}, {Strauss}, {Turner}, {Suzuki}, {Miyazaki}, {Muramatsu}, {Yamamoto},
  {Endo}, {Ezaki}, {Ito}, {Kawaguchi}, {Sofuku}, {Taniike}, {Akutsu}, {Dojo},
  {Kasumi}, {Matsuda}, {Imoto}, {Miwa}, {Suzuki}, {Takeshi}, and
  {Yokota}}]{Miyazaki18}
{Miyazaki}, S., {Komiyama}, Y., {Kawanomoto}, S., {et~al.} 2018, \pasj, 70, S1

\bibitem[{{Nagai} {et~al.}(2007){Nagai}, {Vikhlinin}, and
  {Kravtsov}}]{nagai/vikhlinin/kravtsov:2007}
{Nagai}, D., {Vikhlinin}, A., \& {Kravtsov}, A.~V. 2007, \apj, 655, 98

\bibitem[{{Nelson} {et~al.}(2014{\natexlab{a}}){Nelson}, {Lau}, and
  {Nagai}}]{nelson/etal:2014b}
{Nelson}, K., {Lau}, E.~T., \& {Nagai}, D. 2014{\natexlab{a}}, \apj, 792, 25

\bibitem[{{Nelson} {et~al.}(2014{\natexlab{b}}){Nelson}, {Lau}, {Nagai},
  {Rudd}, and {Yu}}]{nelson/etal:2014a}
{Nelson}, K., {Lau}, E.~T., {Nagai}, D., {Rudd}, D.~H., \& {Yu}, L.
  2014{\natexlab{b}}, \apj, 782, 107

\bibitem[{{Oguri} {et~al.}(2018){Oguri}, {Miyazaki}, {Hikage}, {Mandelbaum},
  {Utsumi}, {Miyatake}, {Takada}, {Armstrong}, {Bosch}, {Komiyama},
  {Leauthaud}, {More}, {Nishizawa}, {Okabe}, and {Tanaka}}]{Oguri18}
{Oguri}, M., {Miyazaki}, S., {Hikage}, C., {et~al.} 2018, \pasj, 70, S26

\bibitem[{{Piffaretti} \& {Valdarnini}(2008){Piffaretti} and
  {Valdarnini}}]{piffaretti/etal:2008}
{Piffaretti}, R., \& {Valdarnini}, R. 2008, \aap, 491, 71

\bibitem[{{Planck Collaboration} {et~al.}(2016{\natexlab{a}}){Planck
  Collaboration}, {Ade}, {Aghanim}, {Arnaud}, {Ashdown}, {Aumont},
  {Baccigalupi}, {Banday}, {Barreiro}, {Bartlett}, and
  et~al.}]{planck2015_cosmo/etal:2016}
{Planck Collaboration}, {Ade}, P.~A.~R., {Aghanim}, N., {et~al.}
  2016{\natexlab{a}}, \aap, 594, A13

\bibitem[{{Planck Collaboration} {et~al.}(2016{\natexlab{b}}){Planck
  Collaboration}, {Aghanim}, {Arnaud}, {Ashdown}, {Aumont}, {Baccigalupi},
  {Banday}, {Barreiro}, {Bartlett}, {Bartolo}, and
  et~al.}]{planck2015_sz/etal:2016}
{Planck Collaboration}, {Aghanim}, N., {Arnaud}, M., {et~al.}
  2016{\natexlab{b}}, \aap, 594, A22

\bibitem[{{Rasia} {et~al.}(2006){Rasia}, {Ettori}, {Moscardini}, {Mazzotta},
  {Borgani}, {Dolag}, {Tormen}, {Cheng}, and {Diaferio}}]{rasia/etal:2006}
{Rasia}, E., {Ettori}, S., {Moscardini}, L., {et~al.} 2006, \mnras, 369, 2013

\bibitem[{{Rasia} {et~al.}(2012){Rasia}, {Meneghetti}, {Martino}, {Borgani},
  {Bonafede}, {Dolag}, {Ettori}, {Fabjan}, {Giocoli}, {Mazzotta}, {Merten},
  {Radovich}, and {Tornatore}}]{rasia/etal:2012}
{Rasia}, E., {Meneghetti}, M., {Martino}, R., {et~al.} 2012, New Journal of
  Physics, 14, 055018

\bibitem[{{Riess} {et~al.}(2019){Riess}, {Casertano}, {Yuan}, {Macri}, and
  {Scolnic}}]{riess/etal:2019}
{Riess}, A.~G., {Casertano}, S., {Yuan}, W., {Macri}, L.~M., \& {Scolnic}, D.
  2019, \apj, 876, 85

\bibitem[{{Riess} {et~al.}(2018{\natexlab{a}}){Riess}, {Casertano}, {Yuan},
  {Macri}, {Bucciarelli}, {Lattanzi}, {MacKenty}, {Bowers}, {Zheng}, and
  {Filippenko}}]{riess/etal:2018b}
{Riess}, A.~G., {Casertano}, S., {Yuan}, W., {et~al.} 2018{\natexlab{a}}, \apj,
  861, 126

\bibitem[{{Riess} {et~al.}(2018{\natexlab{b}}){Riess}, {Casertano}, {Yuan},
  {Macri}, {Anderson}, {MacKenty}, {Bowers}, {Clubb}, {Filippenko}, and
  {Jones}}]{riess/etal:2018a}
---. 2018{\natexlab{b}}, \apj, 855, 136

\bibitem[{{Sakr} {et~al.}(2018){Sakr}, {Ili{\'c}}, {Blanchard}, {Bittar}, and
  {Farah}}]{sakr/etal:2018}
{Sakr}, Z., {Ili{\'c}}, S., {Blanchard}, A., {Bittar}, J., \& {Farah}, W. 2018,
  \aap, 620, A78

\bibitem[{{Salvati} {et~al.}(2018){Salvati}, {Douspis}, and
  {Aghanim}}]{salvati/etal:2018}
{Salvati}, L., {Douspis}, M., \& {Aghanim}, N. 2018, \aap, 614, A13

\bibitem[{{Salvati} {et~al.}(2019){Salvati}, {Douspis}, {Ritz}, {Aghanim}, and
  {Babul}}]{salvati/etal:2019}
{Salvati}, L., {Douspis}, M., {Ritz}, A., {Aghanim}, N., \& {Babul}, A. 2019,
  \aap, 626, A27

\bibitem[{{Shi} \& {Komatsu}(2014){Shi} and {Komatsu}}]{shi/komatsu:2014}
{Shi}, X., \& {Komatsu}, E. 2014, \mnras, 442, 521

\bibitem[{{Shi} {et~al.}(2016){Shi}, {Komatsu}, {Nagai}, and
  {Lau}}]{shi/etal:2016}
{Shi}, X., {Komatsu}, E., {Nagai}, D., \& {Lau}, E.~T. 2016, \mnras, 455, 2936

\bibitem[{{Shi} {et~al.}(2015){Shi}, {Komatsu}, {Nelson}, and
  {Nagai}}]{shi/etal:2015}
{Shi}, X., {Komatsu}, E., {Nelson}, K., \& {Nagai}, D. 2015, \mnras, 448, 1020

\bibitem[{{Stern} {et~al.}(2019){Stern}, {Dietrich}, {Bocquet}, {Applegate},
  {Mohr}, {Bridle}, {Carrasco Kind}, {Gruen}, {Jarvis}, {Kacprzak}, {Saro},
  {Sheldon}, {Troxel}, {Zuntz}, {Benson}, {Capasso}, {Chiu}, {Desai},
  {Rapetti}, {Reichardt}, {Saliwanchik}, {Schrabback}, {Gupta}, {Abbott},
  {Abdalla}, {Avila}, {Bertin}, {Brooks}, {Burke}, {Carnero Rosell},
  {Carretero}, {Castander}, {D'Andrea}, {da Costa}, {Davis}, {De Vicente},
  {Diehl}, {Doel}, {Estrada}, {Evrard}, {Flaugher}, {Fosalba}, {Frieman},
  {Garc{\'\i}a-Bellido}, {Gaztanaga}, {Gruendl}, {Gschwend}, {Gutierrez},
  {Hollowood}, {Jeltema}, {Kirk}, {Kuehn}, {Kuropatkin}, {Lahav}, {Lima},
  {Maia}, {March}, {Melchior}, {Menanteau}, {Miquel}, {Plazas}, {Romer},
  {Sanchez}, {Schindler}, {Schubnell}, {Sevilla-Noarbe}, {Smith}, {Smith},
  {Sobreira}, {Suchyta}, {Swanson}, {Tarle}, and {Walker}}]{stern/etal:2019}
{Stern}, C., {Dietrich}, J.~P., {Bocquet}, S., {et~al.} 2019, \mnras, 485, 69

\bibitem[{Sunyaev \& Zeldovich(1972)Sunyaev and
  Zeldovich}]{sunyaev/zeldovich:1972}
Sunyaev, R.~A., \& Zeldovich, {\relax Ya}.~B. 1972, Comments Astrophys. Space
  Phys., 4, 173

\bibitem[{{Tinker} {et~al.}(2008){Tinker}, {Kravtsov}, {Klypin}, {Abazajian},
  {Warren}, {Yepes}, {Gottl{\"o}ber}, and {Holz}}]{tinker/etal:2008}
{Tinker}, J., {Kravtsov}, A.~V., {Klypin}, A., {et~al.} 2008, \apj, 688, 709

\bibitem[{{Troxel} {et~al.}(2018){Troxel}, {MacCrann}, {Zuntz}, {Eifler},
  {Krause}, {Dodelson}, {Gruen}, {Blazek}, {Friedrich}, and
  {Samuroff}}]{troxel/etal:2018}
{Troxel}, M.~A., {MacCrann}, N., {Zuntz}, J., {et~al.} 2018, \prd, 98, 043528

\bibitem[{{Villaescusa-Navarro} {et~al.}(2014){Villaescusa-Navarro}, {Marulli},
  {Viel}, {Branchini}, {Castorina}, {Sefusatti}, and
  {Saito}}]{villaescusa-Navarro/etal:2014}
{Villaescusa-Navarro}, F., {Marulli}, F., {Viel}, M., {et~al.} 2014, Journal of
  Cosmology and Astro-Particle Physics, 2014, 011

\end{thebibliography}

\end{document}